\documentclass{article}

\usepackage{arxiv}

\usepackage[utf8]{inputenc} 
\usepackage[T1]{fontenc}    
\usepackage{hyperref}       
\usepackage{url}            
\usepackage{booktabs}       
\usepackage{amsfonts}       
\usepackage{nicefrac}       
\usepackage{microtype}      
\usepackage{lipsum}
\usepackage{graphicx}

\usepackage{epstopdf, epsfig}
\usepackage{amsmath}
\usepackage{xcolor}
\usepackage{bm}

\newcommand{\vc}{\mathbf{c}}
\newcommand{\vs}{\mathbf{s}}
\newcommand{\vg}{\mathbf{g}}

\newcommand{\vv}{\mathbf{v}}
\newcommand{\vw}{\mathbf{w}}
\newcommand{\vp}{\mathbf{p}}
\newcommand{\vb}{\mathbf{b}}
\newcommand{\uu}{\mathbf{u}}
\newcommand{\xx}{\mathbf{x}}
\newcommand{\yy}{\mathbf{y}}
\newcommand{\hh}{\mathbf{h}}
\newcommand{\FF}{\mathbf{F}}

\newcommand{\QQ}{\mathbf{Q}}
\newcommand{\II}{\mathbf{I}}

\newcommand{\MS}{\mathbf{S}}
\newcommand{\WW}{\mathbf{W}}

\DeclareMathOperator*{\argmin}{argmin}

\newtheorem{lemma}{Lemma}

\newtheorem{theorem}{Theorem}

\title{Vortex criteria can be objectivized by unsteadiness minimization}

\author{
 Holger Theisel \\
  Department of Computer Science\\
  Otto-von-Guericke University of Magdeburg\\
  Magdeburg, Germany \\
  \texttt{theisel@ovgu.de} \\
   \And
 Markus Hadwiger \\
  Visual Computing Center\\
  King Abdullah University of Science and Technology \\
  Thuwal, 23955-6900, Saudi Arabia \\
  \texttt{markus.hadwiger@kaust.edu.sa} \\
  \And
 Peter Rautek \\
  Visual Computing Center\\
  King Abdullah University of Science and Technology \\
  Thuwal, 23955-6900, Saudi Arabia \\
  \texttt{peter.rautek@kaust.edu.sa} \\
  \And
 Thomas Theu{\ss}l \\
  Visualization Core Lab\\
  King Abdullah University of Science and Technology \\
  Thuwal, 23955-6900, Saudi Arabia \\
  \texttt{thomas.theussl@kaust.edu.sa} \\
  \And
 Tobias G{\"u}nther \\
  Department of Computer Science\\
  Friedrich-Alexander-University of Erlangen-Nuremberg\\
  Erlangen, Germany \\
  \texttt{tobias.guenther@fau.de} \\
}

\begin{document}
\maketitle
\begin{abstract}
Reference frame optimization is a generic framework to calculate a spatially-varying observer field that views an unsteady fluid flow in a reference frame that is as-steady-as-possible. In this paper, we show that the optimized vector field is objective, i.e., it is independent of the initial Euclidean transformation of the observer. To check objectivity, the optimized velocity vectors and the coordinates in which they are defined must both be connected by an Euclidean transformation. In this paper we show that a recent publication \cite{Haller20:can} applied this definition incorrectly, falsely concluding that reference frame optimizations are not objective. Further, we prove the objectivity of the variational formulation of the reference frame optimization proposed in \cite{Haller20:can}, and discuss how the variational formulation relates to recent local and global optimization approaches to unsteadiness minimization.
\end{abstract}


\section{Introduction}
In fluid mechanics, an important property of vortex detectors is whether their corresponding vortex criteria are \emph{objective}, i.e., indifferent to the reference frame in which they are computed. This has been recognized, for example, in the seminal work by Haller \cite{haller2005}, and in a large body of subsequent work.
Non-objectivity implies that different observers, undergoing time-dependent relative rigid motion, might obtain different results for the same conceptual criteria. For example, vortex core lines might be detected at different spatial locations or not be detected at all. This is a major drawback of non-objective methods that often corresponds to the fact that the detected features lack a clear physical meaning, or cannot occur physically at all, as pointed out by many authors, from the early work of Haller~\cite{haller2005} until a recent analysis~\cite{Haller20:can}.
With this motivation in mind, a variety of vortex criteria have been specifically designed to be objective by definition, i.e., the associated method can directly be proven to be indifferent to the motion of the input reference frame, and all observers agree on the result of the evaluated criteria. Usually, these combine (1) \emph{new} proposed criteria, and (2) a direct proof, specific to these criteria, that the proposed method is in fact objective.

One might ask the question whether it is possible to come up with a \emph{generic} way of \emph{``objectivizing''} existing, by themselves non-objective, vortex criteria. If such an approach were successful, it would automatically convert criteria that were not defined in a way that makes them objective by definition, into somehow \emph{equivalent but objective criteria}.
In the literature, three different approaches can be found that aim to objectivize existing vortex criteria: (1) replace the vorticity tensor by the relative spin tensor, (2) replace the vorticity tensor by the spin-deviation tensor, (3) observation in a reference frame that is  as-steady-as-possible.
Approaches of categories (1) and (2) are based on the observation that the vorticity tensor $\WW = (\nabla\vv - \nabla\vv^T)/2$ is not objective, which is, however, frequently used in many vortex definitions~\cite{Jeong95,Hunt87}, cf.~G\"unther and Theisel~\cite{Guenther18:STAR} for a recent review.
Thus, Drouot and Lucius~\cite{Drouot76} and Astarita~\cite{Astarita79} utilized that the strain-rate tensor $\MS = (\nabla\vv + \nabla\vv^T)/2$ is objective, by observing the vorticity tensor $\WW$ in the eigenvector basis of the strain-rate tensor, leading to the relative spin tensor, which can be used as a replacement for the vorticity tensor in all existing vortex criteria.
More recently, Liu et al.~\cite{Liu19:SpinDeviation} utilized that the vorticity can be made objective by subtracting the average vorticity from a local neighborhood, cf.~Haller~\cite{haller2016}, leading to a relative spin tensor, which can be used as building block in existing vortex criteria.
Haller~\cite{Haller20:can} pointed out that the replacement of $\WW$ by the relative spin tensor invalidates the arguments used in the derivation of existing vortex criteria, unless a corresponding frame change is performed.
The first approach of category (3) was proposed by
G\"unther et al.~\cite{Gunther17}, who presented a generic approach by searching for spatially-varying reference frames in which the flow appears as-steady-as-possible. This was motivated by the fact that for steady velocity fields vortices are easier to define. The need for spatially-varying reference frames was pointed out by Lugt~\cite{Lugt79} and Perry and Chong~\cite{Perry94}, who observed that features moving at different speed need differently moving reference frames to make them steady. This idea has created an amount of follow-up research: G\"unther and Theisel~\cite{Guenther:2020:TVCG} consider locally affine frame changes, Baeza Rojo and G\"unther~\cite{Rojo20} incorporate general non-rigid frame changes described by a local Taylor expansion. G\"unther and Theisel~\cite{Guenther:2019:VIS}  extend the approach to inertial flows. Hadwiger et al.~\cite{Hadwiger19} describe frame changes by formulating their derivatives as approximate Killing vector fields. Rautek et al.~\cite{Rautek21} extend this to flows on general two-manifolds.

In a recent paper, Haller \cite{Haller20:can} formulates a variational problem similar to the ones solved in \cite{Gunther17,Hadwiger19,Rojo20,Guenther:2020:TVCG,Guenther:2019:VIS,Rautek21}. Then, Haller~\cite{Haller20:can} attempts to prove that the solution of this variational problem is not objective. From this, Haller~\cite{Haller20:can} concludes that the approaches of G\"unther et al.~\cite{Gunther17}, Hadwiger et al.~\cite{Hadwiger19}, Baeza Rojo and G\"unther~\cite{Rojo20}, and G\"unther and Theisel~\cite{Guenther:2020:TVCG} are not objective either, which is contrary to what is claimed and proven in the respective papers. In addition, Haller~\cite{Haller20:can} claims "physical and mathematical inconsistencies" in \cite{Gunther17,Hadwiger19,Rojo20,Guenther:2020:TVCG}.

The recent paper \cite{Haller20:can} has great importance to research in visualization. If the statements in \cite{Haller20:can} were correct, a significant amount of recent research in visualization would be wrong, including \cite{Gunther17,Hadwiger19,Rojo20,Guenther:2020:TVCG,Guenther:2019:VIS,Rautek21}. Because of this, a careful analysis of the statements in \cite{Haller20:can} is necessary.

In this paper, we make the following contributions:
\begin{itemize}
\item 
We 
show that the proof of the non-objectiveness of  the variational problem by Haller~\cite{Haller20:can}  is not correct because \cite{Haller20:can} applies the definition of objectivity in an  incorrect way.  In particular, we show that \cite{Haller20:can} attempts to check objectivity of vector fields in wrong reference frames.
\item
We show that the variational problem in \cite{Haller20:can} gives objective solutions if it is considered in the correct frame, i.e.,  the optimized velocity field is observed in frames consistent with the motion of the coordinates. The correct objectivized velocity field has the closed form Eq.~\eqref{eq_thm_eq}. 
\item
We show that existing objectivization approaches \cite{Gunther17,Hadwiger19} incorporate the transformation to this correct frame and are therefore objective. 
\item
We show that the claimed mathematical inconsistencies are suitable and necessary boundary conditions  to solve the minimization problem. 
\end{itemize}
We emphasize that the standard definition of objectivity used in continuum mechanics and visualization, as given by Truesdell and Noll~\cite{truesdell_book}, is purely mathematical in nature. Its immediate physical meaning is only that if a method is objective, different physical observers come to the same conclusions, for example regarding the location of a vortex. This is true for all generic ``objectivization'' approaches~\cite{Gunther17,Hadwiger19,Rojo20,Guenther:2020:TVCG,Guenther:2019:VIS,Rautek21}. In contrast to this, however, the argumentation of Haller~\cite{Haller20:can} goes partially beyond objectivity, and in part argues against objectivization of vortex criteria with additional physical considerations. These considerations, however, do not invalidate the objectivity of generic objectivization approaches, and, most importantly, they go beyond the standard definition of objectivity. In this paper, we therefore focus purely on objectivization with the standard meaning of
objectivity, and show that the corresponding mathematical proof given by Haller~\cite{Haller20:can} is incorrect, and that such an objectivization is indeed possible.

\section{The variational problem by Haller \cite{Haller20:can}  }
We set out to show that a reference frame optimization towards an as-steady-as-possible vector field is objective. For this, we demonstrate that the result of the reference frame optimization for the same vector field observed in two different frames is connected through the objectivity condition if observed in the appropriate coordinates.

\subsection{Definition of Objectivity}
We begin with recapitulating the common definition of objectivity, in particular for vector fields. Let $\vw(\xx,t)$ be a vector field observed in a frame (coordinate system) $\FF$. Further, let $\widetilde{\vw}(\yy,t)$ be the observation of $\vw(\xx,t)$ under the Euclidean frame change 
\begin{equation}
\label{eq_def_objectivity_1}
\xx = \QQ(t) \yy + \vb(t),
\end{equation}
where $\QQ(t)$ is a time-dependent rotation tensor and $\vb(t)$ a time-dependent translation vector. 
 Then, $\vw(\xx,t)$ is objective if, cf.~Truesdell and Noll~\cite{truesdell_book}:
\begin{equation}
\label{eq_def_objectivity_2}
\widetilde{\vw}(\yy,t) = \QQ^T(t) \; \vw(\xx,t).
\end{equation}
Note that for the objectivity condition in Eq.~\eqref{eq_def_objectivity_2} to hold, the two vector fields ${\vw}$ and $\widetilde{\vw}$ must be observed in coordinates $\xx$ and $\yy$, respectively, which are connected by Eq.~\eqref{eq_def_objectivity_1}. Furthermore, condition \eqref{eq_def_objectivity_2} must hold for every possible Euclidean transformation \eqref{eq_def_objectivity_1}.

\subsection{Reference Frame Optimization}
A reference frame optimization as in G\"unther et al.~\cite{Gunther17}, Baeza Rojo and G\"unther~\cite{Rojo20}, Hadwiger et al.~\cite{Hadwiger19}, and Rautek et al.~\cite{Rautek21}, aims to view a given vector field in a new reference frame in which the flow becomes as-steady-as-possible, as explained in the following.
To setup the notation, we are given a velocity field $\vv(\xx,t)$ that is observed in the reference frame $\FF$. Further, we assume that $\vv(\xx,t)$ is given in the domain
 $(U,T)$, with $U$ being a simply connected spatial domain and $T=[t_{min},t_{max} ]$ being a time interval.  Observing $\vv(\xx,t)$ in a new reference frame $\FF_*$ given by
\begin{equation}
\label{eq_FF_to_FFstar}
\FF \to \FF_*:\;\; \xx \to \xx_* = \vg(\xx,t)
\end{equation}
results in the observed velocity field
\begin{equation}
\label{eq_eq_v_transformation_1}
\vv_*(\xx_*,t) = \partial_\xx \vg (\vg^{-1}(\xx_*,t),t) \; \vv(\vg^{-1}(\xx_*,t),t) 
+ \partial_t \vg(\vg^{-1}(\xx_*,t),t).
\end{equation}
Here,  $\vg(\xx,t)$ is a diffeomorphism describing a generalized frame change. Note that $\vv_*(\xx_*,t)$ is defined in the domain $(\vg(U,T),T)$, i.e., $\xx_* \in \vg(U,T)$. 

Haller~\cite{Haller20:can} describes a variational problem, which measures the unsteadiness for the transformed flow. This is calculated by integrating the transformed time partial derivatives of $\vv_*(\xx_*,t)$
\begin{equation}
\label{eq_define_Jg}
J(\vg) = \int_{U \times T} \| \partial_t \vv_*(\xx_*,t)\|^2 dV
\end{equation}
which serves as objective for the optimal frame change $\hat{\vg}$ resulting in the minimizer
\begin{equation}
\label{eq_minimizer_ghat}
\hat{\vg} = \argmin_{\vg \in C^2(U\times T)} J(\vg).
\end{equation}
\begin{figure*}%
    \centering%
    \includegraphics[width=\linewidth]{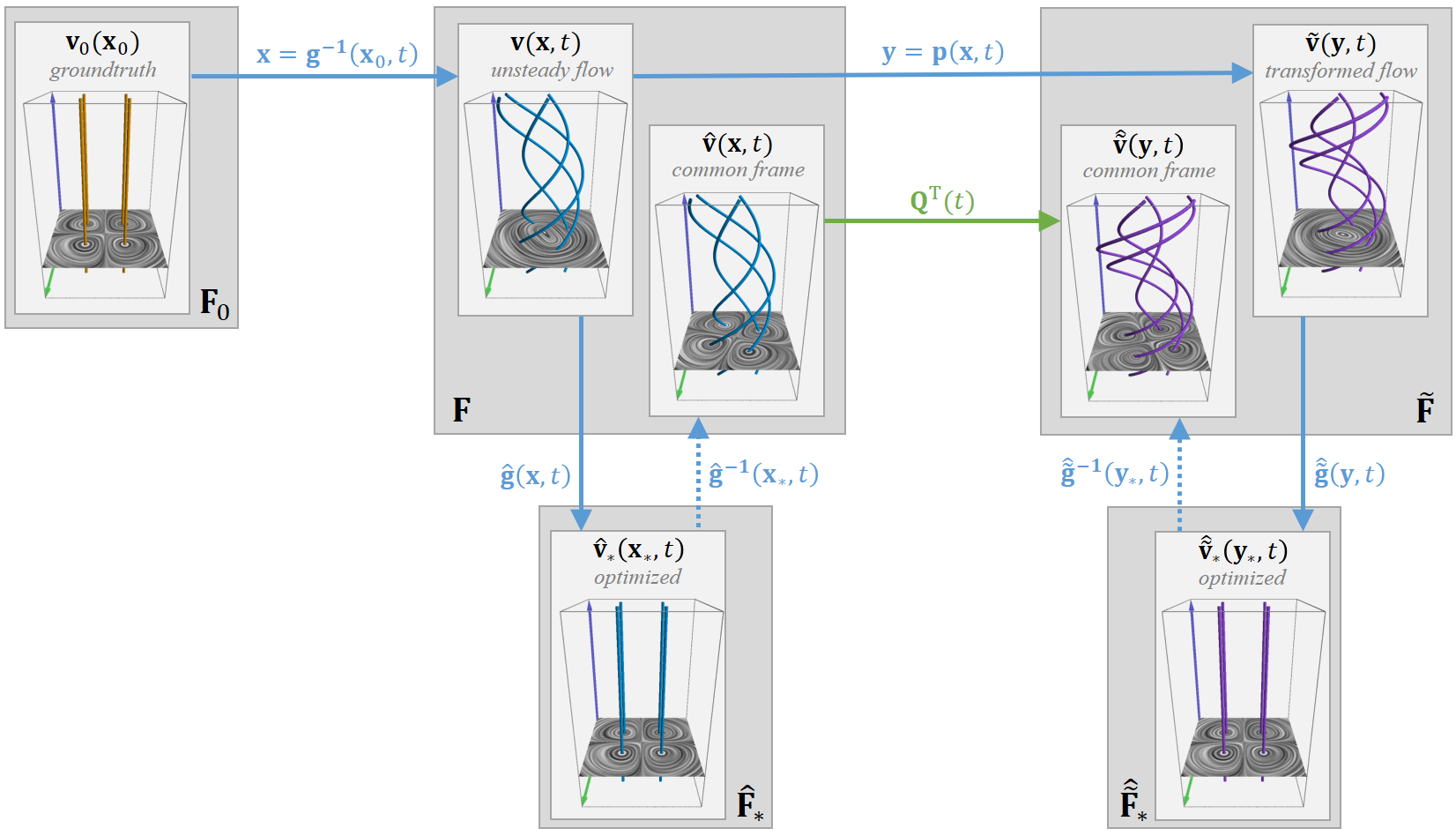}%
    \caption[.]{Illustration of reference frame transformations along with an example in 2D space-time $(x,y,t)$. A steady flow $\vv_0(\xx_0)$ is observed in two arbitrary reference frames connected by $\vp(\xx,t)$, resulting in the observed unsteady flows $\vv(\xx,t)$ and $\widetilde\vv(\yy,t)$. Note that their planar line integral convolution (LIC) slices do not show the correct vortex structures (shown as lines), since they depict streamlines and not pathlines of the unsteady flow. Assuming a unique solution, a reference frame optimization will result for both in the same steady flow. Deforming the optimized flows $\hat\vv_*(\xx_*,t)$ and $\hat{\widetilde\vv}_*(\yy_*,t)$ to coordinates $\xx$ and $\yy$ (dashed arrows), results in the vector fields $\hat\vv(\xx,t)$ and $\hat{\widetilde\vv}(\yy,t)$, which are connected by the rotation in $\vp(\xx,t)$. Note that these deformed vector fields show vortex structures in the LIC slices at the vortex locations. In this figure, we used
    $\vv_0(x,y)=\begin{pmatrix}-x(2y^2-1)\\ ~~~y(2x^2-1) \end{pmatrix}e^{-x^2-y^2}$ and 
    $\vg^{-1}(\xx,t) = \begin{pmatrix} \cos(t) & -\sin(t) \\ \sin(t) & ~~~\cos(t)\end{pmatrix} \xx + \begin{pmatrix}t/10 \\ 0\end{pmatrix}$ and
    $\vp(\xx,t) = \begin{pmatrix} \cos(\frac{t}{2}) & -\sin(\frac{t}{2}) \\ \sin(\frac{t}{2}) & ~~~\cos(\frac{t}{2})\end{pmatrix} \xx + \begin{pmatrix}t/5 \\ 0\end{pmatrix}$, with the fields being shown in the domain $[-2,2]^2\times[0,2\pi]$. In $\vv_0(x,y)$, vortex centers are located at $(\pm 1/\sqrt{2}, \pm 1/\sqrt{2})$.
    }%
    \label{fig:illustration-frames}%
\end{figure*}%
\begin{figure*}[t]
    \centering
    \includegraphics[width=\linewidth]{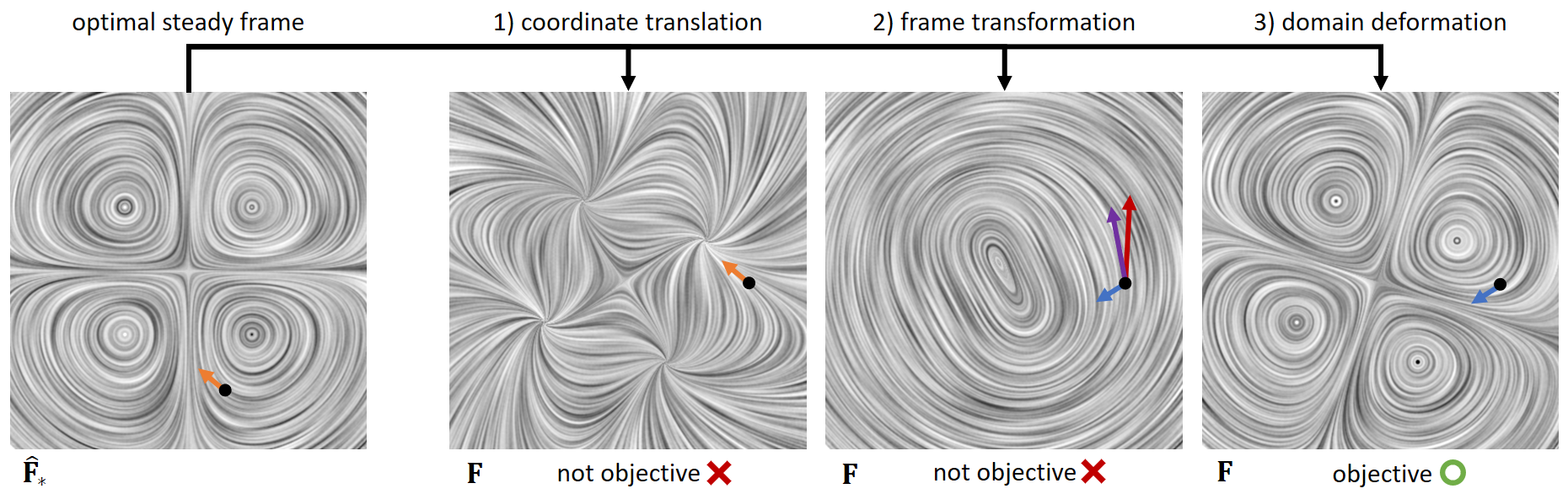}%
    \caption{In principle, there are three different transformations that can take us from $\widehat\FF_* \to \FF$. 1) The coordinate translation simply translates the vector of the steady frame (orange) in $\widehat\FF_*$ via $\hat\vg^{-1}$ to $\FF$. While critical point locations are preserved, the flow around critical points is not physically meaningful and dependent on the observer motion. 2) The inverse reference frame transformation takes the steady flow back to the original unsteady flow $\vv(\xx,t)$ (purple vector). Note that the purple vector is the superposition of the deformed steady frame (blue vector) and the inverse motion of the observer (red vector). Due to the latter, the result is dependent on the observer motion and therefore not objective. 3) The inverse domain deformation (dashed arrow in Fig.~\ref{fig:illustration-frames}) results in an objective vector field (blue vector). Note that unlike 1) and 2), the approach 3) is objective. We used the same vector field and transformations as in Fig.~\ref{fig:illustration-frames}, and show the time slice $t=2\pi/5$.}%
    \label{fig:transformations}%
\end{figure*}

\subsection{Proof of Non-Objectivity in \cite{Haller20:can}}
Haller~\cite{Haller20:can} claims that the resulting optimal velocity field
\begin{equation}
\hat{\vv}_*(\xx_*,t) = \partial_\xx \hat{\vg} (\hat{\vg}^{-1}(\xx_*,t),t) \; \vv(\hat{\vg}^{-1}(\xx_*,t),t) 
+ \partial_t \hat{\vg}(\hat{\vg}^{-1}(\xx_*,t),t)
\end{equation}
from 
the solution of the variational problem (\ref{eq_define_Jg}),
(\ref{eq_minimizer_ghat}) can never be objective.
The setup for the proof of non-objectivity in Haller~\cite{Haller20:can} is illustrated in Figure~\ref{fig:illustration-frames} and starts with an arbitrary steady vector field $\vv_0(\xx_0)$ observed in a reference frame $\FF_0$, and a diffeomorphism $\vg(\xx,t)$, which describes a frame change. From this, a time-dependent velocity field $\vv(\xx,t)$ in the reference frame $\FF$ is obtained by observing $\vv_0(\xx_0)$ under the inverse frame change
\begin{equation}
\label{eq_define_v_from_steadyfield}
\FF_0 \to \FF:\;\; \xx_0 \to \xx =  \vg^{-1}(\xx_0,t).
\end{equation}
This means that we construct an unsteady vector field $\vv(\xx,t)$ for which the reference frame transformation with $\vg(\xx,t)=\xx_0$ (our ground truth) takes us back to the steady flow $\vv_0(\xx_0)$.
For $\vv(\xx,t)$, we search the optimal frame change $\hat{\vg}$ minimizing  (\ref{eq_minimizer_ghat}). Applying $\hat{\vg}$ to $\vv(\xx,t)$ results in the optimal observed velocity field $\hat{\vv}_*(\xx_*,t)$ in the frame $\hat{\FF}_*$ by
\begin{equation}
\FF \to \hat{\FF}_*:\;\; \xx \to \xx_* = \hat{\vg}(\xx,t).
\end{equation}
To test the objectivity of the reference frame optimization, we need to observe the unsteady velocity field $\vv(\xx,t)$ (observed in $\FF$) in an arbitrary reference frame $\widetilde{\FF}$ relative to it, and apply the reference frame optimization there, too.
Applying a frame change according to Eq.~\eqref{eq_def_objectivity_1} gives the new coordinates $\yy$ with
\begin{equation}
\label{eq_def_objectivity_3}
\FF \to \widetilde{\FF}:\;\ \xx \to \yy = \QQ^T(t)( \xx - \vb(t)) = \vp(\xx,t).
\end{equation}
Thus, observing the field $\vv(\xx,t)$ in  a moving Euclidean frame $\widetilde{\FF}$ as given by $\vp(\xx,t)$ in 
(\ref{eq_def_objectivity_3}) results in the field
\begin{align}
\widetilde{\vv}(\yy,t) &= \partial_\xx \vp (\vp^{-1}(\yy,t)) \; \vv(\vp^{-1}(\yy,t),t) + \partial_t \vp(\vp^{-1}(\yy,t))\\
&= \QQ^T(t) \left( \vv(\QQ(t)\yy + \vb(t) ,t) - \dot{\QQ}(t) \yy - \dot{\vb}(t)   \right).  
\label{eq_define_vtilde}
\end{align}
Also for $\widetilde{\vv}(\yy,t)$, we search the optimal frame change $\hat{\widetilde{\vg}}$ minimizing  (\ref{eq_minimizer_ghat}). Applying $\hat{\widetilde{\vg}}$ to $\widetilde{\vv}(\yy,t)$ results in the optimal observed velocity field $\hat{\widetilde{\vv}}_*({\yy}_*,t)$ in the frame $\hat{\widetilde{\FF}}_*$ by
\begin{equation}
\widetilde{\FF} \to \hat{\widetilde{\FF}}_*:\;\; \yy \to {\yy}_* = \hat{\widetilde{\vg}}(\yy,t).
\end{equation}
From the particular construction of $\vv(\xx,t)$ by (\ref{eq_define_v_from_steadyfield}), both $\hat{\vg}(\xx,t)$ and $\hat{\widetilde{\vg}}(\yy,t)$ have closed-form solutions:
\begin{equation}
\hat{\vg}(\xx,t)=\vg(\xx,t) \;\;\;,\;\;\;
\hat{\widetilde{\vg}}(\yy,t) = \vg( \vp^{-1}(\yy,t),t)
\end{equation}
which means that they both reach the ground truth steady vector field:
\begin{align}
\hat{\vv}_*(\xx_*,t) &= 
 \vv_0( \xx_* ) \\
\hat{\widetilde{\vv}}_*({\yy}_*,t) &= 
\vv_0( \yy_* )
\end{align}
Using the rotation $\QQ^T(t)$ from Eq.~\eqref{eq_def_objectivity_3}, which connects $\xx$ and $\yy$, in an attempt to test the objectivity condition in Eq.~\eqref{eq_def_objectivity_2} therefore gives an inequality
\begin{equation}
\label{eq_countercondition}
\hat{\widetilde{\vv}}_*(\yy_*,t) \neq \QQ^T(t) \; \hat{\vv}_*(\xx_*,t)
\end{equation}
because $\hat{\vv}_*$ and $\hat{\widetilde{\vv}}_*$ are steady and $\QQ^T(t)$ is truly time-dependent. 
From  (\ref{eq_countercondition}), Haller~\cite{Haller20:can} concluded non-objectivity because condition (\ref{eq_def_objectivity_2}) of the objectivity condition is not fulfilled.

This conclusion, however, is not correct,  
because the prerequisite (\ref{eq_def_objectivity_1}) for checking the objectivity condition (\ref{eq_def_objectivity_2}) is not fulfilled in the first place, i.e., $\xx_* $ and $\yy_*$ are not connected by the rotation $\QQ(t)$ via Eq.~\eqref{eq_def_objectivity_3}, i.e.:
\begin{equation}
\label{eq_countercondition2}
\xx_* \neq \QQ(t) \yy_* + \vb(t).
\end{equation}
Keep in mind that the common objectivity definition has the form "if (\ref{eq_def_objectivity_1}) then (\ref{eq_def_objectivity_2})". To check for   objectivity (\ref{eq_def_objectivity_2}), we must compare vector fields in frames with a relative motion (\ref{eq_def_objectivity_1}) to each other. This, however, is not the case for the reference frames $\hat{\FF}_*$, $\hat{\widetilde{\FF}}_*$ in which $\hat{\vv}_*$ and $\hat{\widetilde{\vv}}_*$ are observed. 
In fact, the relation of $\hat{\FF}_*$ and $\hat{\widetilde{\FF}}_*$ is
\begin{equation}
\hat{\FF}_* \to \hat{\widetilde{\FF}}_*:\;\ \xx_* \to \yy_* = \xx_* \neq  \vp( \xx_*,t),
\end{equation}
i.e., $\hat{\FF}_*$ and $\hat{\widetilde{\FF}}_*$ are identical. Since in this setting  (\ref{eq_def_objectivity_1}) does not apply, we cannot make any conclusions about objectivity or non-objectivity.
\subsection{Proof of Objectivity}

To correctly check for objectivity, we have to transform $\hat{\vv}_*(\xx_*,t)$ from $\hat{\FF}_*$ to $\FF$, resulting in $\hat\vv(\xx,t)$, and we have to transform $\hat{\widetilde{\vv}}_*(\yy_*,t)$ from $\hat{\widetilde{\FF}}_*$ to $\widetilde{\FF}$, resulting in
$\hat{\widetilde\vv}(\yy,t)$.
This way, both vector fields are in coordinates $\xx$ and $\yy$, respectively, which are indeed connected by $\QQ(t)$ in Eq.~\eqref{eq_def_objectivity_3}. 

For these transformations 
$\hat{\vv}_*(\xx_*,t) \to
\hat\vv(\xx,t)$ 
and 
$\hat{\widetilde{\vv}}_*(\yy_*,t) \to
\hat{\widetilde\vv}(\yy,t)$, several options are possible and require a discussion. In order to support Haller's \cite{Haller20:can} general statement ("solution of (\ref{eq_define_Jg}),
(\ref{eq_minimizer_ghat}) can never be objective"), it is necessary to show that all transformations 
$\hat{\vv}_*(\xx_*,t) \to
\hat\vv(\xx,t)$ 
and 
$\hat{\widetilde{\vv}}_*(\yy_*,t) \to
\hat{\widetilde\vv}(\yy,t)$
lead to non-objective vector fields 
$\hat{\vv}$ and $\hat{\widetilde{\vv}}$. Further, to show that an existing frame optimization approach is non-objective, one has to identify which transformations $\hat{\vv}_*(\xx_*,t) \to
\hat\vv(\xx,t)$ 
and 
$\hat{\widetilde{\vv}}_*(\yy_*,t) \to
\hat{\widetilde\vv}(\yy,t)$ are used, and for them non-objectivity has to be shown.

In principle, there are three ways how such transformations 
$\hat{\vv}_*(\xx_*,t) \to
\hat\vv(\xx,t)$ 
and 
$\hat{\widetilde{\vv}}_*(\yy_*,t) \to
\hat{\widetilde\vv}(\yy,t)$
could be conceived, which are illustrated in Fig.~\ref{fig:transformations} for the LIC slice shown in Fig.~\ref{fig:illustration-frames}:
\begin{enumerate}
    \item Simply translate coordinates, giving $\hat\vv(\hat\vg(\xx,t),t)$ and $\hat{\widetilde\vv}(\hat{\widetilde\vg}(\yy,t),t)$, which does not account for rotations of the observer. This is neither physically meaningful nor objective.
    \label{option-1}
    \item Apply an inverse reference frame transformation according to Eq.~\eqref{eq_eq_v_transformation_1} using the inverse of $\hat\vg(\xx,t)$ and $\hat{\widetilde\vg}(\yy,t)$, respectively (reverse of solid arrows in Fig.~\ref{fig:illustration-frames}). This simply results in the original unsteady flows $\vv(\xx,t)$ and $\widetilde\vv(\yy,t)$. The resulting flow is physically observable, but the approach is not objective.%
    \label{option-2}
    \item Use an inverse domain deformation only (dashed arrows in Fig.~\ref{fig:illustration-frames}). This is objective and results in a derived vector field that unveils flow features of the original time-dependent flow. This transformation is later introduced in 
    (\ref{eq_vhat1}), (\ref{eq_vhat2}).
    \label{option-3}
\end{enumerate}
In the following we formally introduce the transformation in \ref{option-3}). Then we prove that this transformation gives objective vector fields $\hat\vv(\xx,t)$ and $\hat{\widetilde\vv}(\yy,t)$. We also show that 
existing local reference frame optimizations~\cite{Gunther17,Guenther:2020:TVCG,Rojo20} are equivalent to  \ref{option-3}).
To explain this further, we first formally introduce the inverse transformations
\begin{eqnarray}
\hat{\FF}_* \to \FF:\;\ \xx_* \to \xx = \hat\hh(\xx_*,t) \\
\hat{\widetilde{\FF}}_* \to \widetilde{\FF}:\;\ \yy_* \to \yy = \hat{\widetilde\hh}(\yy_*,t)
\end{eqnarray}
with $\hat\hh = \hat\vg^{-1}$ and $\hat{\widetilde\hh}=\hat{\widetilde\vg}^{-1}$, i.e.,
\begin{eqnarray}
\label{eq_define_h}
\hat\hh(\hat\vg(\xx,t),t)=\xx \;\;\;,\;\;\;
\hat{\widetilde\hh}(\hat{\widetilde\vg}(\yy,t),t)=\yy.
\end{eqnarray}
Computing the spatial gradients in (\ref{eq_define_h}) gives
\begin{eqnarray}
\label{eq_define_h_2}
\partial_\xx \hat\hh(\xx_*,t) \; \partial_\xx \hat\vg(\xx,t) =
\partial_\xx \hat{\widetilde\hh}(\yy_*,t) \; \partial_\xx \hat{\widetilde\vg}(\yy,t) =
\II
\end{eqnarray}
Note that for transforming $\hat{\vv}_*(\xx_*,t) \to \hat{\vv}(\xx,t)$ and 
$\hat{\widetilde{\vv}}_*(\yy_*,t) \to \hat{\widetilde{\vv}}(\yy,t)$, we spatially deform the vector field to the appropriate coordinates rather than applying a reference frame transformation, which would result in the original unsteady flows.
Deforming the optimized vector field places the flow structures that are observed in the optimal frame at their locations in the original frame, revealing for example vortex cores as critical points or lines with swirling motion around them.
The deformation of the optimal flows is:
\begin{eqnarray}
\label{eq_vhat1}
 \hat{\vv}(\xx,t) &=& \partial_\xx \hat\hh(\xx_*,t) \;  \hat{\vv}_*(\xx_*,t)\\
\label{eq_vhat2} 
 \hat{\widetilde{\vv}}(\yy,t) &=& \partial_\xx \hat{\widetilde{\hh}}(\yy_*,t) \;  \hat{\widetilde{\vv}}_*(\yy_*,t).
 \end{eqnarray}
Inserting the transformations from the given unsteady flows to the optimal flows, i.e.,
\begin{eqnarray}
\hat{\vv}_*(\xx_*,t) &=& \partial_\xx \hat\vg(\xx,t) \; \vv(\xx,t)
+ \partial_t\hat\vg(\xx,t)\\
\hat{\widetilde{\vv}}_*(\yy_*,t) &=& \partial_\xx \hat{\widetilde\vg}(\yy,t) \; {\widetilde\vv}(\yy,t)
+ \partial_t\hat{\widetilde\vg}(\yy,t)
\end{eqnarray}
into (\ref{eq_vhat1}), (\ref{eq_vhat2}), while using (\ref{eq_define_h_2}) gives the deformed flows $\hat{\vv}(\xx,t)$ and $\hat{\widetilde{\vv}}(\yy,t)$ in terms of the original unsteady flows $\vv(\xx,t)$ and $\widetilde{\vv}(\yy,t)$:
 \begin{eqnarray}
 \hat{\vv}(\xx,t) 
  &=& \vv(\xx,t) + (\partial_\xx \hat{\vg}(\xx,t))^{-1} \; \partial_t\hat{\vg}(\xx,t)  \label{objective-field}\\
  \hat{\widetilde{\vv}}(\yy,t) 
   &=& \widetilde{\vv}(\yy,t) + (\partial_\xx \hat{\widetilde{\vg}}(\yy,t))^{-1} \; \partial_t\hat{\widetilde{\vg}}(\yy,t).
 \end{eqnarray}
Since now $\hat\vv(\xx,t)$ and  $\hat{\widetilde\vv}(\yy,t)$ are in the frames $\FF$, $\widetilde{\FF}$ fulfilling (\ref{eq_def_objectivity_1}), we can check for objectivity using (\ref{eq_def_objectivity_2}). We formulate
\begin{theorem}
\label{eq_thm_1}
Given is a velocity field $\vv(\xx,t)$. If the variational problem (\ref{eq_define_Jg}), (\ref{eq_minimizer_ghat}) has a unique minimizer $\hat\vg$ (up to a steady Euclidean frame change), then the field 
\begin{equation}
\label{eq_thm_eq}
\hat{\vv} = \vv + (\partial_\xx \hat{\vg})^{-1} \; \partial_t\hat{\vg}
\end{equation}
in Eq.~\eqref{objective-field} is objective.
\end{theorem}
This theorem is the main theoretical result of this paper.
For proving this, we start with 
\begin{lemma}
\label{lemma1}
Given is a vector field $\vv(\xx,t)$ and its observation  
$\widetilde{\vv}(\yy,t)$ under the Euclidean frame change (\ref{eq_def_objectivity_3}). Further, let 
$\vv_*(\xx_*,t)$ be the observation of $\vv$ under the frame change (\ref{eq_FF_to_FFstar}), i.e., $\vv_*$ is given by (\ref{eq_eq_v_transformation_1}). Finally, let $\widetilde{\vv}_*(\yy_*,t)$ be the observation of $\widetilde{\vv}$ under the observation frame 
\begin{equation}
\label{eq_tildeFF_tildeFFstar}
    \widetilde{\FF} \to \widetilde{\FF}_*:  \yy \to  \yy_* = \widetilde{\vg}(\yy,t)  =    \vg(\vp^{-1}(\yy,t),t).
\end{equation}
Then 
\begin{eqnarray}
\label{eq_lemma1}
\widetilde{\vv}_*(\yy_*,t) = \vv_*(\xx_*,t).
\end{eqnarray}
\end{lemma}
Proof: Both $\vv_*$ and $\widetilde{\vv}_*$ are obtained by observing $\vv$ in the moving reference frames $\FF_*$, $\widetilde{\FF}_*$, respectively. $\FF_*$ is given in (\ref{eq_FF_to_FFstar}). (\ref{eq_def_objectivity_3}) and (\ref{eq_tildeFF_tildeFFstar}) 
 give for $\widetilde{\FF}_*$:
\begin{align}
\label{eq_tildeFF_tilde}
   \FF \to \widetilde{\FF} \to \widetilde{\FF}_* : \xx \to \yy=\vp(\xx,t) \to  \yy_* \\
   \yy_* = \widetilde{\vg}(\yy,t)  =    \vg(\vp^{-1}(\yy,t),t) = \vg(\xx,t).
\end{align}
This and (\ref{eq_FF_to_FFstar}) give $\widetilde{\FF}_* = \FF_*$ which proves (\ref{eq_lemma1}).
Note that (\ref{eq_lemma1}) holds for a general vector field $\vv(\xx,t)$ and is not limited to fields constructed from steady vector fields via reference frame transformation, as done with Eq.~\eqref{eq_define_v_from_steadyfield}.

From Lemma \ref{lemma1} it follows for $\vg(\xx,t)$ and $\widetilde{\vg}(\yy,t)$, connected via Eq.~\eqref{eq_tildeFF_tildeFFstar}, that
\begin{equation}
\label{eq_J_Jtilde}
J(\vg) = \widetilde{J}(\widetilde{\vg})
\end{equation}
where $J(\vg)$ is given in (\ref{eq_define_Jg}) and
\begin{equation}
\label{eq_define_Jg_tilde}
\widetilde{J}(\widetilde{\vg}) = \int_{U \times T} \| \partial_t \widetilde{\vv}_*(\yy_*,t)\|^2 dV. 
\end{equation}
From this follows:
\begin{lemma}
\label{lemma2}
If (\ref{eq_define_Jg}) has the unique minimizer $\hat\vg$, i.e., (\ref{eq_minimizer_ghat}) holds, 
then $\hat{\widetilde{\vg}}(\yy,t) = \hat\vg( \vp^{-1}(\yy,t),t)$ is the unique mimimizer of
(\ref{eq_define_Jg_tilde}), i.e., 
\begin{equation}
\label{eq_minimizer_ghat_tilde}
\hat{\widetilde{\vg}} = \argmin_{\widetilde{\vg} \in C^2(U\times T)} \widetilde{J}(\widetilde{\vg}).
\end{equation}
\end{lemma}
Finally, from this, (\ref{eq_lemma1}), and 
\begin{eqnarray}
\partial_\xx \hat{\widetilde\vg}(\yy,t) =
\partial_\xx  \vg(\xx,t)  \;
\partial_\xx (\vp^{-1}(\yy,t))
=\partial_\xx  \vg(\xx,t)  \; \QQ(t)
\end{eqnarray}
follows
\begin{eqnarray}
 \hat{\widetilde{\vv}}(\yy,t) &=& \partial_\xx \hat{\widetilde{\hh}}(\yy_*,t) \;  \hat{\widetilde{\vv}}_*(\yy_*,t)
\\ 
&=& (\partial_\xx \hat{\widetilde\vg}(\yy,t))^{-1}
\;  \hat{\widetilde{\vv}}_*(\yy_*,t)
\\ 
&=& \QQ^T(t) \;  
(\partial_\xx  \hat\vg(\xx,t))^{-1}
\;  \hat{\widetilde{\vv}}_*(\yy_*,t)
\\ 
&=& \QQ^T(t) \;  
\partial_\xx \hat\hh(\xx_*,t) \;  \hat{\vv}_*(\xx_*,t)
\\
&=& \QQ^T(t) \; \hat\vv(\xx,t)
\end{eqnarray}
which proves Theorem \ref{eq_thm_1}.

\subsubsection*{Remark:}

It is important to note that the optimal vector field 
$\hat{\vv}$ in 
\eqref{eq_thm_eq} is objective but not observable from $\vv$ in the sense that there is a general frame change $\vg$ such that $\hat{\vv}$ is the observation of $\vv$ under $\vg$. Because of this, care has to be taken which vortex extractors are applied to $\hat{\vv}$. While local measures (such as the $Q$ criterion) applied to $\hat{\vv}$ generally give good results (as done, e.g., in G\"unther et al.~\cite{Gunther17}), Lagrangian approaches (based on an integration of $\hat{\vv}$) are not advisable since a trajectory in  $\hat{\vv}$ does not have a physical meaning.

\subsection{Uniqueness Considerations}

In the following, we show that a solution of 
(\ref{eq_define_Jg}),
(\ref{eq_minimizer_ghat})  can be unique only up to a steady (time-independent)  Euclidean  frame change. Let
 $ \vs(\xx_*) =  \MS \; \xx_* + \vs_c $
be a steady Euclidean frame change, i.e., $\MS$ is a rotation matrix and $\vs_c$ is a translation vector. We define the frame change
\begin{equation}
\label{eq_uniquenesstransform}
\FF \to \FF_{**}: \xx \to \xx_{**} = \vg_{\vs}(\xx,t) = \vs(\vg(\xx,t))
\end{equation}
where $g(\xx,t)$ takes us from $\xx$ to $\xx_*$ and $s(\xx_*)$ takes us from $\xx_*$ to $\xx_{**}$.
Then, we  get the partial derivatives
\begin{equation}
\label{eq_uniquenesstransform2}
\partial_\xx \vg_{\vs}(\xx,t) = \MS \; \partial_\xx \vg(\xx,t) \;\;\;,\;\;\;
\partial_t \vg_{\vs}(\xx,t) = \MS \; \partial_t \vg(\xx,t).
\end{equation}
Thus, a reference frame transformation with (\ref{eq_uniquenesstransform})
transforms $\vv$ to
\begin{align}
\vv_{**}(\xx_{**},t) =&~ \partial_\xx \vg_{\vs} (\vg_{\vs}^{-1}(\xx_{**},t),t) \; \label{eq:steady-0} \vv(\vg_{\vs}^{-1}(\xx_{**},t),t) \\
&+ \partial_t \vg_{\vs}(\vg_{\vs}^{-1}(\xx_{**},t),t) \label{eq:steady-1}\\
=&~ \partial_\xx \vg_{\vs} ( \xx   ,t) \; \vv(  \xx  ,t) + \partial_t \vg_{\vs}(  \xx   ,t) \label{eq:steady-2}\\
=&~ \MS\; \partial_\xx \vg( \xx   ,t) \; \vv(  \xx  ,t) + \MS\; \partial_t \vg(  \xx   ,t) \label{eq:steady-3}\\
=&~ \MS \;\vv_{*}(\xx_{*},t). \label{eq:steady-4}
\end{align}
where the step from Eq.~\eqref{eq:steady-0}--\eqref{eq:steady-1} to Eq.~\eqref{eq:steady-2} applies the inverse transformation of Eq.~\eqref{eq_uniquenesstransform} to the coordinates, Eq.~\eqref{eq:steady-2} to Eq.~\eqref{eq:steady-3} applies Eq.~\eqref{eq_uniquenesstransform2}, and Eq.~\eqref{eq:steady-3} to Eq.~\eqref{eq:steady-4} transforms the velocity field to coordinates $\xx_*$, resulting in the factored out matrix $\MS$.
This gives $\partial_t \vv_{**}(\xx_{**},t) = \MS \; \partial_t \vv_{*}(\xx_{*},t)$, resulting in 
\begin{equation}
\label{eq_uniqenessproblem}
J(\vg) = J(\vg_{\vs}).
\end{equation}
This has the following meaning: if $\hat\vg(\xx,t)$ is a minimizer of (\ref{eq_define_Jg}),
(\ref{eq_minimizer_ghat}), then $\hat\vg_{\vs}(\xx,t) = \vs(\hat\vg(\xx,t))$ is a minimizer as well. Because of this, a proper boundary condition has to "pick" a particular $\vs$ to ensure a unique solution of the variational problem. Fortunately, this picking does not influence the final objective velocity field in the frame $\FF$:
\begin{align}
\label{eq_uniqenessproblem_solution}
&~\vv(\xx,t) + (\partial_\xx \hat{\vg}(\xx,t))^{-1} \; \partial_t\hat{\vg}(\xx,t) \\
=&~ \vv(\xx,t) + (\partial_\xx \hat{\vg}_{\vs}(\xx,t))^{-1} \; \partial_t \hat{\vg}_{\vs}(\xx,t).
\end{align}
(\ref{eq_uniqenessproblem_solution}) follows directly from (\ref{eq_uniquenesstransform2}):
\begin{equation}
(\partial_\xx \hat{\vg}_{\vs})^{-1} \; \partial_t \hat{\vg}_{{\vs}} =
(\partial_\xx \hat{\vg})^{-1} \; \MS^{-1} \; \MS \; \partial_t\hat{\vg} =
(\partial_\xx \hat{\vg})^{-1} \; \partial_t\hat{\vg}.
\end{equation}

\section{Relation to Existing Approaches}

In order to come up with practical solutions for the variational problem (\ref{eq_define_Jg}),
(\ref{eq_minimizer_ghat}), several further design decisions are necessary:
\begin{itemize}
\item
Choice of the domain: Since it is not generally possible to find a ``perfect'' frame change for an arbitrary unsteady vector field (i.e., a frame where $\hat\vv_*$ becomes perfectly steady) for the whole domain $(U,T)$, certain subsets of $(U,T)$ may be considered instead.
\item
Limitations to subclasses of $\vg$: The space of all considered frame changes $\vg$ can be limited, e.g., to the space of all Euclidean frame changes.
\item
Boundary conditions: depending on the settings above, proper boundary conditions are necessary to ensure a unique solution of (\ref{eq_define_Jg}),
(\ref{eq_minimizer_ghat}). In particular, the uniqueness problem (\ref{eq_uniqenessproblem}) needs to be addressed by the boundary conditions.
\item
Non-linearity: the problem (\ref{eq_define_Jg}),
(\ref{eq_minimizer_ghat}) is a non-linear PDE, making a numerical solution challenging. Existing approaches \cite{Gunther17},\cite{Hadwiger19} manage to transform it to a linear PDE before solving it. 

\end{itemize}
In the following, we discuss how existing approaches for unsteadiness minimization are related to the variational problem (\ref{eq_define_Jg}),
(\ref{eq_minimizer_ghat}). We restrict ourselves to the approaches of G\"unther et al.~\cite{Gunther17} and Hadwiger et al.~\cite{Hadwiger19}, respectively, because the other existing approaches build upon them.

\subsection{The approach by Günther et al.~\cite{Gunther17}}

Günther et al.~\cite{Gunther17} consider only Euclidean frame changes~$\vg$.
Further, the approach in \cite{Gunther17} does not directly solve (\ref{eq_define_Jg}),
(\ref{eq_minimizer_ghat}). In particular, it does not solve 
(\ref{eq_define_Jg}),
(\ref{eq_minimizer_ghat}) 
under the additional condition 
\begin{equation}
\label{eq_rebuttal_3}
\vg(\xx,t)=\xx  \;\;\;,\;\;\; \partial_\xx  \vg(\xx,t)=\II
\end{equation}
as claimed by Haller~\cite{Haller20:can}. Instead, Günther et al.~\cite{Gunther17} solve a similar problem as (\ref{eq_define_Jg}),
(\ref{eq_minimizer_ghat}) for each point individually by assuming an individual neighborhood for each point. Let $\varepsilon, \gamma>0$ be small constants, let $U_\varepsilon(\vc)$ be the spatial $\varepsilon$-neighborhood around $\vc$, and let $T_\gamma(\tau)$ be the time-neighborhood around $\tau$. Further, we assume  $\varepsilon, \gamma$ small enough to fulfill $(U_\varepsilon(\vc),T_\gamma(\tau)) \subseteq (U,T)$. Then, Günther et al.~\cite{Gunther17} solve an individual variational problem for each space-time location $(\vc,\tau)$ by
\begin{eqnarray}
\label{eq_define_Jg_individual}
J(\vg_{(\vc,\tau)}) &=& \int_{U_\varepsilon(\vc) \times T_\gamma(\tau)} \| \partial_t \vv_*(\xx_*,t)\|^2 dV 
\\
\label{eq_minimizer_ghat_individual}
\hat{\vg}_{(\vc,\tau)} &=& \argmin_{\vg_{(\vc,\tau)} \in C^2(U_\varepsilon(\vc) \times T_\gamma(\tau))} J(\vg_{(\vc,\tau)}).
\end{eqnarray}
This means that $\hat{\vg}_{(\vc,\tau)}(\xx,t)$  is the optimal frame change when considering only the neighborhood around $(\vc,\tau)$. From this, Günther et al.~\cite{Gunther17} consider the parameter-dependent vector field 
\begin{equation}
\hat\vv'(\xx,t;\,\vc,\tau) = \vv(\xx,t) + (\partial_\xx \; \hat{\vg}_{(\vc,\tau)}(\xx,t))^{-1} \;
(\partial_t \; \hat{\vg}_{(\vc,\tau)}(\xx,t))
\end{equation}
from which the final objective vector field
\begin{equation}
\hat\vv(\xx,t) = \hat\vv'(\xx,t;\,\xx,t)
\end{equation}
is derived.
To compute $\hat\vv'(\xx,t;\,\vc,\tau)$, we need to compute $\hat{\vg}_{(\vc,\tau)}$ for every $(\vc,\tau)$.
For this, certain boundary conditions for the uniqueness problem (\ref{eq_uniqenessproblem}) are necessary. Günther et al.~\cite{Gunther17} use the conditions
\begin{equation}
\label{eq_boundaryconditions}
\hat{\vg}_{(\vc,\tau)}(\xx,t)_{|t=\tau}=\xx  \;\;\;,\;\;\;
\partial_\xx \; \hat{\vg}_{(\vc,\tau)}(\xx,t)_{|t=\tau} = \II.
\end{equation}
which sets conditions only in a single time slice, namely at the observation time $t=\tau$, i.e., the reference frame is free to deform locally in the space-time neighborhood.
This particular choice of the boundary conditions has the advantage that for each $(\vc,\tau)$
\begin{equation}
\label{eq_omitbacktransformation}
      \hat\vv_*(\vc,\tau)    = \hat\vv(\vc,\tau)
\end{equation}
i.e., it is sufficient to compute $\hat\vv_*$ without the final transformation (\ref{eq_vhat1}) from $\FF_*$ to $\FF$. (Note that (\ref{eq_omitbacktransformation}) directly follows from (\ref{eq_vhat1}) and (\ref{eq_boundaryconditions})).
Finally Günther et al.~\cite{Gunther17} solve the problem for $\gamma \to 0$, making it possible to completely represent $\vg$ by a Taylor approximation. With this, the solution of  
(\ref{eq_define_Jg_individual})
(\ref{eq_minimizer_ghat_individual}) turns out to be a quadratic problem for each $(\vc,\tau)$ with the $t$-derivatives of $\vg_{(\vc,\tau)}(\xx,t)$ as unknowns.
In follow-up work, the frame change $\vg$ received further degrees of freedom~\cite{Guenther:2020:TVCG,Rojo20}.

\subsection{The approach by Hadwiger et al.~\cite{Hadwiger19}}

Hadwiger et al.~\cite{Hadwiger19} take another approach to solve (\ref{eq_define_Jg}),
(\ref{eq_minimizer_ghat}). Instead of searching for optimal frame changes $\vg$, they directly solve for the vector fields
\begin{equation}
\label{eq_killing1}
\uu = - (\partial_\xx \vg)^{-1} \vg_t,
\end{equation}
where the right-hand side of \eqref{eq_killing1}
appears in the right-hand side of
 (\ref{eq_thm_eq}). This has the advantage that the uniqueness problem  
(\ref{eq_uniqenessproblem}) does not have to be addressed because of 
(\ref{eq_uniqenessproblem_solution}). In particular, Hadwiger et al.~\cite{Hadwiger19} search for approximate Killing vector fields, which is justified by the following
\begin{lemma}
If $\vg$ is an Euclidean frame change, then $\uu$ is a Killing vector field.
\end{lemma}
Further, the relation to the variational problem 
(\ref{eq_define_Jg}), (\ref{eq_minimizer_ghat}) is given by
\begin{align}
\label{eq_killing2}
\partial_t \vv_*(\xx_*,t) =&~  
\partial_\xx \vg(\xx,t) \; L_{\uu} (\vv-\uu)(\xx,t)\\
\nonumber
=&~
\partial_\xx \vg(\xx,t) \;
\left(
\partial_\xx\vv(\xx,t)\; \uu(\xx,t) + \partial_t\vv(\xx,t) \right.\\
&- \left.\partial_\xx\uu(\xx,t)\; \vv(\xx,t)
- \partial_t\uu(\xx,t)
\right),
\end{align}
where $L_{\uu}$ denotes the time-dependent Lie derivative. In this way, Hadwiger et al.~\cite{Hadwiger19} solve the variational problem
\begin{eqnarray}
\label{eq_hadwiger1}
J(\uu) &=& \int_{U \times T} \| L_{\uu} (\vv-\uu)(\xx,t)\|^2 dV  \\
\hat{\uu} &=& \argmin_{\uu \in C^1(U\times T)} J(\uu)\\
\label{eq_hadwiger3}
\hat{\vv} &=& \vv - \hat{\uu}.
\end{eqnarray}
If the search space is restricted to perfect Euclidean frame changes 
$\vg$ (i.e., exact Killing fields $\uu$), Eqs.~\eqref{eq_hadwiger1}--\eqref{eq_hadwiger3} are  -- due to \eqref{eq_killing1}--\eqref{eq_killing2} --  identical to \eqref{eq_define_Jg}--\eqref{eq_minimizer_ghat} but have a number of practical advantages: \eqref{eq_hadwiger1}--\eqref{eq_hadwiger3} is linear in the unknown $\uu$, and the uniqueness problem (\ref{eq_uniqenessproblem}) does not have to be addressed.
Haller~\cite{Haller20:can} further claims the problems "not accounting for the $\xx$-dependence of the initial conditions of the flow of their proposed observer vector field", and "frame-change formulas for rotating observers that do not account for the rotation of the observer" in \cite{Hadwiger19}. We disagree: Both the $\xx$-dependence as well as the rotation are encoded in the vector field $\uu$, including the corresponding frame change formulas given in Sec.~6.3 of Hadwiger et al.~\cite{Hadwiger19}.

\section{Conclusions}
Nowadays, it is generally agreed upon that vortex criteria should be independent of the chosen reference frame, in particular invariant to Euclidean transformations, which is referred to as objectivity. Many of the commonly-used vortex definitions such as the $\lambda_2$- and $Q$-criterion do not enjoy this mathematical property. To this date, three generic approaches have been proposed to alter these definitions into an objective counterpart, including the replacement of the vorticity tensor with the relative-spin or spin-deviation tensors, or by finding spatially-varying reference frames in which the flow becomes as-steady-as-possible. The latter not only enables the analysis of unsteady flow by means of techniques developed for steady flows, it also makes every existing vortex measure objective. In his recent paper, Haller~\cite{Haller20:can} systematically analyzed these approaches, formulated the reference frame optimization as variational problem, and incorrectly concluded that the optimization is not objective. In this paper, we showed that \cite{Haller20:can} applied the objectivity definition incorrectly by comparing optimized vector fields in the wrong coordinates. In fact, both the velocity vectors of the fields, as well as the coordinates in which they are defined must obey the Euclidean transformation. Hence, we demonstrate that the objectivization via reference frame optimization is in fact objective, and we discuss how the variational problem relates to the local optimization approaches of G\"unther et al.~\cite{Gunther17} and Baeza Rojo and G\"unther~\cite{Rojo20}, and the global optimization of Hadwiger et al.~\cite{Hadwiger19}, which also applies analogously to the recent approach by Rautek et al.~\cite{Rautek21}. We believe that reference frame optimization is a promising device for unsteady vector field analysis, including not only vortices but also other flow features as well as topological elements. 


\bibliographystyle{unsrt}  


\end{document}